# ACEnet: Anatomical Context-Encoding Network for Neuroanatomy Segmentation

Yuemeng Li, Hongming Li, and Yong Fan

Center for Biomedical Image Computing and Analytics and the Department of Radiology, the Perelman School of Medicine at the University of Pennsylvania, Philadelphia, PA 19104 USA

***Abstract***: Segmentation of brain structures from magnetic resonance (MR) scans plays an important role in the quantification of brain morphology. Since 3D deep learning models suffer from high computational cost, 2D deep learning methods are favored for their computational efficiency. However, existing 2D deep learning methods are not equipped to effectively capture 3D spatial contextual information that is needed to achieve accurate brain structure segmentation. In order to overcome this limitation, we develop an Anatomical Context-Encoding Network (ACEnet) to incorporate 3D spatial and anatomical contexts in 2D convolutional neural networks (CNNs) for efficient and accurate segmentation of brain structures from MR scans, consisting of 1) an anatomical context encoding module to incorporate anatomical information in 2D CNNs and 2) a spatial context encoding module to integrate 3D image information in 2D CNNs. In addition, a skull stripping module is adopted to guide the 2D CNNs to attend to the brain. Extensive experiments on three benchmark datasets have demonstrated that our method achieves promising performance compared with state-of-the-art alternative methods for brain structure segmentation in terms of both computational efficiency and segmentation accuracy.
***Keywords:*** Convolutional Neural Networks; context encoding; image segmentation; attention

## INTRODUCTION

Deep learning methods have achieved huge success in a variety of image segmentation studies, including brain structure segmentation from magnetic resonance (MR) scans (Brosch et al., 2016; Chen et al., 2017; Chen et al., 2018; Dai et al., 2019; Huo et al., 2019; Kamnitsas et al., 2017; Lafferty et al., 2001; Li et al., 2017; Moeskops et al., 2016; Wachinger et al., 2018; Zhang et al., 2018; Zhang et al., 2015; Zhao et al., 2017; Zheng et al., 2015). Previous studies on the brain structure segmentation have favored volumetric segmentation based on 3D convolutional neural networks (CNNs) (Brosch et al., 2016; Dai et al., 2019; Huo et al., 2019; Kamnitsas et al., 2017; Li et al., 2017; Moeskops et al., 2016; Wachinger et al., 2018; Zhang et al., 2015). These methods typically build deep learning models on overlapped 3D image patches. In particular, DeepNAT was proposed to predict segmentation labels of 3D image patches under a hierarchical classification and multi-task learning setting (Wachinger et al., 2018); a 3D whole brain segmentation method was developed to segment the brain structures using spatially localized atlas network tiles (SLANT) (Huo et al., 2019); and a transfer learning method was developed to segment the brain structures by learning from partial annotations (Dai et al., 2019). Although these 3D segmentation methods have achieved promising segmentation performance, they are computationally expensive for both model training and inference, and their applicability is potentially hampered by the memory limitation of typical graphics processing units (GPUs).

In order to improve the computational efficiency of deep learning models for the brain image segmentation, a variety of deep learning methods have been developed for segmenting 2D image slices of 3D MRI brain images (Roy et al., 2019; Roy et al., 2017; Roy et al., 2018), in addition to quantized 3D neural networks (Paschali et al., 2019). Particularly, QuickNAT (Roy et al., 2019) was proposed to segment 2D brain image slices in multiple views (Coronal, Axial, Sagittal) using a modified U-Net framework (Ronneberger et al., 2015) with densely connected blocks (Huang et al., 2017). Furthermore, a modified version was developed to improve its performance (Roy et al., 2018) with a joint spatial-wise and channel-wise Squeeze-and-Excitation (SE) module to fuse both spatial and channel information within local receptive fields (Hu et al., 2018). These 2D segmentation methods could segment a whole brain image in ~20 seconds on a typical GPU. However, the 2D segmentation methods ignore intrinsic 3D contextual information of 3D brain MR images, which could potentially improve the segmentation performance if properly utilized.

Most deep learning-based brain structure segmentation methods focus on segmentation of coarse-grained brain structures, and it remains largely unknown if they work well for segmenting the MRI brain images into fine-grained structures. Whereas the fine-grained brain structure segmentation could provide richer neuroanatomy information than a coarse-grain brain structure segmentation in neuroimaging studies of brain development, aging, and brain diseases (Li et al., 2019; Pomponio et al., 2019), it is more challenging as the fine-grained structures are relatively small and with similar image appearances, especially for the 2D segmentation methods that do not utilize 3D contextual information.

To achieve fast and accurate segmentation of fine-grained brain structures from MR scans, we develop a deep neural network for segmenting 2D slices of MR scans by integrating 3D spatial and anatomical contexts in 2D CNNs, inspired by the success of deep learning with contextual information for image segmentation (Chen et al., 2017; Chen et al., 2018; Zhang et al., 2018; Zhao et al., 2017; Zhao et al., 2018; Zheng et al., 2015). Particularly, anatomical context is encoded in 2D CNNs through an attention module with a global anatomy classification supervision and 3D spatial context is encoded in 2D multi-channel input of spatially consecutive image slices. Additionally, the segmentation network also integrates a skull stripping auxiliary task to guide the network to focus on the brain structures. The method has been compared with state-of-the-art competing deep learning methods in terms of computational efficiency and segmentation accuracy based on 3 public datasets, including 2012 Multi-Atlas Labelling Challenge (MALC) dataset (Landman and Warfield, 2012), Mindboggle-101 dataset (Klein and Tourville, 2012), and Schizophrenia Bulletin (SchizBull) 2008 dataset (Kennedy et al., 2012). Based on these datasets, we directly compared our method with Skip-DeconvNet (SD-Net) (Roy et al., 2017), 2D Unet (Ronneberger et al., 2015), QuickNAT V2 (Roy et al., 2018), and 3D Unet (Çiçek et al., 2016), with a focus on methods built upon 2D CNNs for computational efficiency. We also reported image segmentation performance of MO-Net (Dai et al., 2019), SLANT (Huo et al., 2019), 3DQuantized-Unet (Paschali et al., 2019), and DeepNAT (Wachinger et al., 2018) that were evaluated on the 2012 MALC dataset with the same training and testing images, except SLANT. Source code of this study is available at https://github.com/ymli39/ACEnet-for-Neuroanatomy-Segmentation.

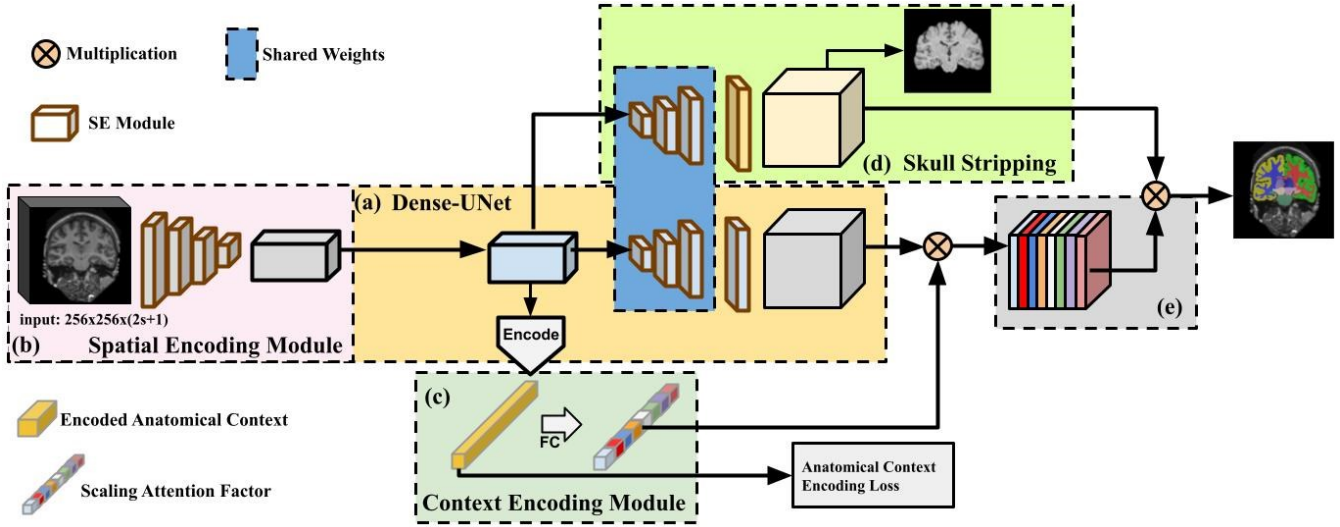

Fig. 1. A schematic flowchart of Anatomy Context-Encoding network. (a) A Dense-UNet backbone. (b) A Spatial Context Encoding Module with a 3D image volume as its input. (c) An Anatomical Context Encoding Module contains a context encoder to capture anatomical context. (d) A Skull Striping Module to enforce the network to specifically focus on the brain. Particularly, the spatial encoding module captures 3D features from the input using 2D CNNs. The context encoder captures anatomical context to highlight brain structure-dependent variation by optimizing an Anatomical Context Encoding Loss. The spatial and anatomical semantics (e) and skull stripping features (d) are fused by an element-wise multiplication operation to generate accurate brain structure segmentation result.

## METHODS

We develop a deep learning method, referred to as Anatomy Context-Encoding network (ACEnet), for segmenting both coarse-grained and fine-grained anatomical structures from brain MR scans. ACEnet is a 2D network for segmenting brain MR scans slice by slice. As illustrated in Fig. 1-(a), ACEnet is built upon a densely connected encoder-decoder backbone, consisting of 1) a 3D spatial context encoding module as shown in Fig. 1-(b) to integrate spatial appearance information using 2D CNNs; 2) an anatomical context encoding module as shown in Fig. 1-(c) to incorporate anatomical information in 2D CNNs with a classification loss of brain structures; and 3) a skull stripping module as shown in Fig. 1-(d) to guide 2D CNNs to attend the brain. Image features learned by these 2D CNNs are finally fused to segment brain structures as illustrated in Fig. 1-(e). In the present study, we focus on image slices in coronal plane. For clarity, we use "3D" to denote input of a stack of multiple 2D slices to 2D CNNs hereafter.

*A. Network Backbone*

The network backbone is an U-Net (Ronneberger et al., 2015) with 4 densely connected blocks for both the encoder and the decoder, as illustrated in Fig. 1-(a). Each dense block contains 2 padded $5 \times 5$ convolutions followed by a $1 \times 1$ convolution layer. Particularly, max-pooling layers are adopted in the encoder blocks and up-sampling layers are adopted in the decoder blocks. Skip connections are adopted between the encoder and the decoder blocks with the same spatial dimensions. To fuse both spatial-wise and channel-wise information within local receptive fields, spatial and channel Squeeze-and-Excitation (sc-SE) (Roy et al., 2018) is applied to each encoder, bottleneck, and decoder dense blocks. The sc-SE is built upon Spatial Squeeze and Channel Excitation (c-SE) (Hu et al., 2018) and Channel Squeeze and Spatial Excitation (s-SE) (Roy et al., 2018) that are fused by a Max-Out operation to effectively learn both spatial-wise and channel-wise information. The c-SE block has a hyper-parameter $r$ that was set to 2 in the present study for all experiments as suggested in (Roy et al., 2018). In this backbone setting, our goal is to learn image features for effective brain structure segmentation.

*B. Spatial Context Encoding Module*

To utilize 3D spatial information of MR scans in ACEnet, 3D image blocks of consecutive image slices are used as input to the spatial context encoding module, as illustrated in Fig. 1-(b). The consecutive image slices are regarded as a stack of 2D images with dimensions of $H \times W \times C$, where $H$ and $W$ are spatial dimensions of the 2D image slices and $C$ is the number of 2D image slices, rather than as a 3D volume with dimensions of $H \times W \times C \times 1$. Therefore, the input to the spatial context encoding module is of the same dimensions as the 2D input. Particularly, we set $C = 2s + 1$, where $s$ is the number of consecutive 2D image slices stacked on top and bottom of the center slice that is the image slice to be segmented. For an image slice without top or bottom adjacent slices, we used the image slice itself as its adjacent slices. Instead of directly implementing a 3D CNN module, which is computationally expensive, the spatial context encoding module acquires intrinsic spatial context information with less computation cost. This module takes the 3D input to the encoder and outputs 2D feature representation with 3D spatial context that is used as input to the anatomical context encoding module (Fig. 1-(c)) and the decoder.

*C. Anatomical Context Encoding Module*

The anatomical context encoding module is developed to integrate global anatomical information in ACEnet. As illustrated in Fig. 1-(c), the output of the network bottleneck is used as input to the anatomical context encoding module, consisting of a convolutional block, referred to as encoding layer, a fully connected layer, and an activation function. The anatomical context encoding module is applied to output of the network bottleneck that contains high level information learned from the data with a reduced dimensionality. The anatomical context is learned through the encoding layer and is then passed through the fully connected layer followed by a sigmoid activation function that detects the presence of specific brain structures in the center slice of the input. Particularly, the detection of the presence of specific brain structures is formulated as a classification problem with an anatomical context encoding loss (ACE-loss) to optimize the network under a direct supervision. It specifically focuses on the brain structures present in the 3D input's center image slice under consideration, rather than all the brain structures to be segmented. The output of the anatomical context encoder is referred to as encoded anatomical context.

To facilitate the semantic segmentation, the encoded anatomical context is utilized to extract the global semantic context represented by a scaling attention factor as shown in Fig. 1. This scaling attention factor, denoted by $\gamma$, is the output of a sigmoid function $\sigma(\cdot)$, i.e., $\gamma = \sigma(We)$, where $W$ is the layer weight and $e$ is the encoded anatomical context. This scaling attention factor provides the network with the global anatomical context to squeeze the intensity ambiguity between brain structures with similar appearances, and to selectively highlight the learned feature maps associated with specific brain structures present in the input of 3D image block's center slice. This scaling factor is also utilized to recalibrate the decoded output, calculated as $Y = X \otimes \gamma$, where $X$ denotes feature maps generated from the decoder and $\otimes$ is a channel-wise multiplication. We refer to this recalibrated output as fused semantics.

*D. Skull Stripping Module*

In order to guide the brain structure segmentation network to focus on the brain structures, rather than non-brain structures such as nose and neck region, we include a skull stripping module as an auxiliary task to extract the brain from MR scans, as illustrated in Fig. 1-(d). The first three decoders of the Skull Stripping Module share the same weight as the model backbone's decoders and only its last decoder block is trained with separate weight parameters to reduce the model complexity. The skull stripping module learns informative features in a supervised manner with a skull stripping loss function. The learned image features are combined with the recalibrated output as illustrated in Fig. 1-(e) to generate the brain structure segmentation labels.

*E. Loss Function*

We use three loss functions to train the network, including (i) a pixel-wise cross-entropy loss $L_{ce}$, (ii) a multi-class Dice loss $L_{dice}$, and (iii) an anatomical context encoding classification loss $L_{sec}$. The pixel-wise cross-entropy loss measures similarity between output segmentation labels and manual labeled ground truth (Shore and Johnson, 1980). Denote the estimated probability of a pixel $x$ belonging to a class $l$ by $p_l(x)$ and its ground truth label by $g_l(x)$, the pixel-wise cross-entropy loss is:

$$L_{ce} = -\sum_x g_l(x) \log(p_l(x)).$$

The multi-class Dice score is often used as an evaluation metric in image segmentation studies. In the present study, we include the multi-class Dice loss function to overcome class-imbalance problem (Roy et al., 2019; Roy et al., 2017), which is formulated as:

$$L_{dice} = -\frac{2\sum_x p_l(x)g_l(x)}{\sum_x p_l^2(x) + \sum_x g_l^2(x)}.$$

The anatomical context encoding loss is used to incorporate anatomical information in 2D CNNs so that the network focuses on specific brain structures present in the input of 3D image block's center slice:

$$L_{sec} = -\frac{1}{C}\sum_{i=1}^{C} y_i \cdot \log(p(y_i)) + (1 - y_i) \cdot \log(1 - p(y_i)),$$

where $C$ is the number of classes of brain structures, $y_i$ is the ground truth that a specific brain structure is present or not in the input of 3D image block's center slice, and $p(y)$ is the predicted probability of the presence of that specific brain structure. This loss is adopted to learn the anatomical context as illustrated in Fig. 1-(c).

Both $L_{ce}$ and $L_{dice}$ loss functions are applied to the skull stripping module for skull stripping as $L_{N_{skull}}$, and fused structural segmentation prediction as $L_{N_{brain}}$. Therefore, the overall loss is formulated as:

$$L_{total} = L_{ce_{skull}} + L_{dice_{skull}} + L_{ce_{brain}} + L_{dice_{brain}} + \lambda L_{sec},$$

where $\lambda = 0.1$ is a weighting factor as suggested in (Zhang et al., 2018).

*F. Implementation Details*

Our 2D CNN network takes a 3D image volume as multiple channels of $256 \times 256 \times (2s + 1)$ as inputs, all in coronal view. We employed a learning rate scheduling "poly" that is updated at each $iter$ step as $lr = baselr \times \left(1 - \frac{iter}{iter_{total}}\right)^{power}$ (Chen et al., 2017), where $baselr$ is the initial learning rate. We set power to 0.9 as suggested in (Zhang et al., 2018). We trained our model in two stages as detailed in ablation studies. In the first stage, we chose an initial learning rate of 0.01 and 0.02 for segmenting coarse-grained structures and fine-grained structures, respectively. In the second stage, we set the initial learning rate to 0.01 for both tasks. Both pre-trained and fine-tuned model were trained for 100 epochs. In both the stages, we utilized the SGD optimizer with a momentum of 0.9 and a weight decay rate of $1 \times 10^{-4}$. We used batch size of 6 to use all available GPU memory of a Titan XP GPU. The dropout rate of 0.1 was applied to each densely connected block (Srivastava et al., 2014). All experiments were performed on a single NVIDIA TITAN XP GPU with 12GB of RAM. It took ~9 seconds to obtain both brain structure segmentation and skull-stripping results from an MRI scans of $256 \times 256 \times 256$ on a NVIDIA TITAN XP GPU.

**Experimental Datasets and settings**

*A. Imaging datasets*

We evaluated our method based on three public datasets with manually labelled coarse-grained or fine-grained brain structures, as detailed following.

(i) **2012 Multi-Atlas Labelling Challenge (MALC)**: This dataset contains MRI T1 scans from 30 subjects with manual annotations for the whole brain, including 27 coarse-grained structures and 134 fine-grained structures (Landman and Warfield, 2012). In studies of segmenting coarse-grained brain structures we focused on all available coarse-grained brain structures, and in studies of segmenting fine-grained brain structures we focused on 133 fine-grained structures following BrainColor protocol (Klein et al., 2010). This challenge dataset also provides a list of 15 training subjects and a list of 15 testing subjects. The same training and testing data setting was used in our experiments to train and evaluate deep learning segmentation models.

Based on the 2012 MALC training scans, we generated an augmented training data set. Particularly, we applied deformable registration to warp the training images and their corresponding segmentation labels to twenty 1.5 T MR images, randomly selected from Alzheimer's Disease Neuroimaging Initiative (ADNI) dataset (Petersen et al., 2010). In total, 300 warped images and segmentation label images were obtained as the augmented training dataset.

(ii) **Mindboggle-101**: This dataset contains MRI T1 scans from 101 healthy subjects with 63 manual annotated brain structures (Klein and Tourville, 2012). In the present study, we randomly split the dataset into training (60%), validation (10%), and test (30%) sets. The best validation model was utilized for testing.

(iii) **Schizophrenia Bulletin (SchizBull) 2008**: This dataset is part of the Child and Adolescent Neuro Development Initiative (CANDI) dataset, consisting of MRI T1 scans from 103 subjects with 32 manual labeled brain structures (Kennedy et al., 2012). In the present study, we randomly split the data into training (60%), validation (10%) and test (30%) set. The best validation model was utilized for testing.

In our experiments, all the images were resampled into an isotropic volume of $1mm^3$ by "mri_convert" of FreeSurfer (Fischl, 2012). No other preprocessing procedures were applied to these images. The binary brain masks obtained from FreeSurfer preprocessing are used as ground truth brain regions for training and evaluation in skull stripping stage. We carried out ablation studies to evaluate how different components of our method contribute to the segmentation based on three benchmark datasets.

*B. Ablation studies*

A baseline of the present study was an improved version of QuickNAT (Roy et al., 2019) with sc-SE blocks (Roy et al., 2018), referred to as QuickNAT V2, which was built upon the same Dense U-Net structure as ACEnet. In the ablation studies, the batch size of different deep learning models was set to use all available GPU memory of a Titan XP GPU.

We first evaluated if the pixel-wise cross-entropy loss with the class weights could improve the segmentation performance, in conjunction with different settings of the anatomical context encoding module and the spatial context encoding module. As proposed in QuickNAT (Roy et al., 2019), frequencies of voxel-wise segmentation labels of the training can be used as class weights in the pixel-wise cross-entropy loss. The class weight $\omega(x)$ of a pixel $x$ is computed as:

$$\omega(x) = \sum_l I(S(x) = l)\frac{median(f)}{f_l} + \omega_0 \cdot \mathbb{I}(|\nabla S(x)| > 0),$$

where $f$ denotes a vector of frequencies of all voxel-wise segmentation labels, $l$ denotes a specific segmentation label, and $f_l$ denotes its frequency in the training data, $\mathbb{I}$ is an indicator function, $S$ is the ground truth segmentation label map, $\nabla$ is 2D gradient operator, and $\omega_0 = \frac{2 \cdot median(f)}{f_l}$.

We then investigated the effectiveness of 1) anatomical context encoding module, 2) spatial context encoding module, and 3) skull stripping module. Particularly, we adopted the anatomical context encoding module in six models with different inputs as well as with and without the skull stripping module on three benchmark datasets. We studied various spatial context learned from inputs of 1) single 2D image slice, 2) a stack of multiple 2D image slices, and 3) two parallel encoders with inputs of single 2D image slice and a stack of 2D image slices respectively, and the two sets of encoded output features were concatenated after their specific bottleneck blocks. We incorporated the spatial context with and without skull stripping module to evaluate how the skull stripping module affects the overall segmentation performance.

To comprehend how the sc-SE blocks modulate image features learned by densely connected blocks of CNNs in ACEnet, we generated feature maps and attention maps for each encoder and each decoder to visualize attention maps and image features before and after the sc-SE blocks (Roy et al., 2018; Schlemper et al., 2018). Since all the input and output image features of the encoders and decoders are multi-channel features, we obtained absolute values of image features averaged out across channels to visualize image features learned by different network blocks. Since the spatial-wise and channel-wise attention blocks integratively modulate the image features and the channel-wise attention is characterized by a vector, we generated spatial-wise attention maps and did not visualize the channel-wise attention vectors.

We investigated how the parameter $s$ in the spatial context encoding module affects the segmentation performance, and we also evaluated networks built with different values of $s$ using the end-to-end training setting with the presence of the anatomical context encoding module and skull stripping module.

Moreover, we investigated the effectiveness of the end-to-end training and two-stage training strategies. For the two-stage setting, we trained our model by utilizing only fused semantics outputs (Fig. 1-(e)) without skull stripping module (Fig. 1-(d)) in the first stage; in the second stage we incorporated the pre-trained weights obtained in the first training stage in the proposed architecture and fine-tuned the whole network with the skull stripping module as an auxiliary task. In this ablation study, the end-to-end model was trained with the same number of total epochs (200 epochs) as the two-stage training strategy

*C. Comparison with state-of-the-art competing methods*

We directly compared our method with state-of-the-art competing deep learning methods on the three datasets with the same model training and test settings, including SD-Net (Roy et al., 2017), 2D U-Net (Ronneberger et al., 2015), QuickNAT V2 (Roy et al., 2018), and 3D U-Net (Çiçek et al., 2016). All these methods were implemented with the same network architectures as reported in their corresponding papers, except that 256 filters were used in the 3D U-Net instead of 1024 for reducing the computational cost.

We also reported image segmentation performance of MO-Net (Dai et al., 2019), SLANT (Huo et al., 2019), 3DQuantized-Unet (Paschali et al., 2019), and DeepNAT (Wachinger et al., 2018) that were evaluated on the 2012 MALC dataset for segmenting either coarse-grained or fine-grained brain structures with the same training and testing images, expect SLANT models which were trained on a larger training dataset.

*D. Quantitative evaluation metrics*

The image segmentation performance was evaluated on the testing data using Dice Score, Jaccard Index, and Hausdorff distance between the ground truth and automatically segmented brain structures (Hao et al., 2014; Zheng et al., 2018). Two-side Wilcoxon signed rank tests were adopted to compare ACEnet and QuickNAT V2 in terms of Dice scores of individual brain structures.

Table 1. Dice scores of ACEnet with different settings and its baselines on the 2012 MALC test data with 27 coarse-grained segmentation structures. √ indicates presence of the entry, s is the consecutive image slices, $E_{context}$: Contextual Encoding Module.

| Inputs S=0 | Inputs S=5 | $E_{context}$ | Class Weight | Batch size | Dice Score |
|---|---|---|---|---|---|
| √ |   |   | √ | 10 | 0.851 |
| √ |   |   |   | 10 | 0.876 |
| √ |   | √ | √ | 8 | 0.870 |
| √ |   | √ |   | 8 | 0.887 |
|   | √ | √ | √ | 6 | 0.867 |
|   | √ | √ |   | 6 | 0.885 |

# RESULTS

*A. Ablation Studies on Loss Function, Anatomical Context, Spatial Context, and Skull Stripping*
ACEnet's backbone is a U-Net architecture, consisting of 4 densely connected blocks for both the encoder and the decoder, the exactly same architecture used in QuickNAT V2 (Roy et al., 2018) and serving as the baseline in our experiments. All encoder, bottleneck, and decoder dense blocks contain the sc-SE module (Roy et al., 2018). Table 1 summarizes segmentation performance for segmenting coarse-grained brain structures on the 2012 MALC testing data obtained by deep learning models with different settings of the loss function, anatomical context encoding module, and spatial context encoding module. The segmentation models built with the pixel-wise cross-entropy loss without the class weights had better performance than their counterparts with the class weights in the pixel-wise cross-entropy loss function for the baseline models (top two rows), the baseline models with anatomical context (middle two rows), and the models with both spatial and anatomical context (bottom two rows). In all following experiments, the pixel-wise cross-entropy loss without the class weights was used. The results summarized in Table 1 also indicated that the anatomical context encoding module improved the segmentation performance for the baseline models.

Fig. 2 shows representative spatial-wise attention maps of the sc-SE blocks and maps of image features before and after modulation by the sc-SE blocks for segmenting coarse-grained brain structures on the 2012 MALC data set. Specifically, image features of a randomly selected image slice (top row) were used as input to densely-connected blocks to generate new image features that were subsequently modulated by the sc-SE blocks (their spatial-wise attention maps are shown on the middle row) to yield modulated image features (bottom row). Although the attention maps had varied spatial patterns at different encoders and decoders, they increased contrasts between background and brain tissues of the feature maps, which subsequently improved the segmentation performance as supported by the quantitative results summarized in Table 1 and Table 2.

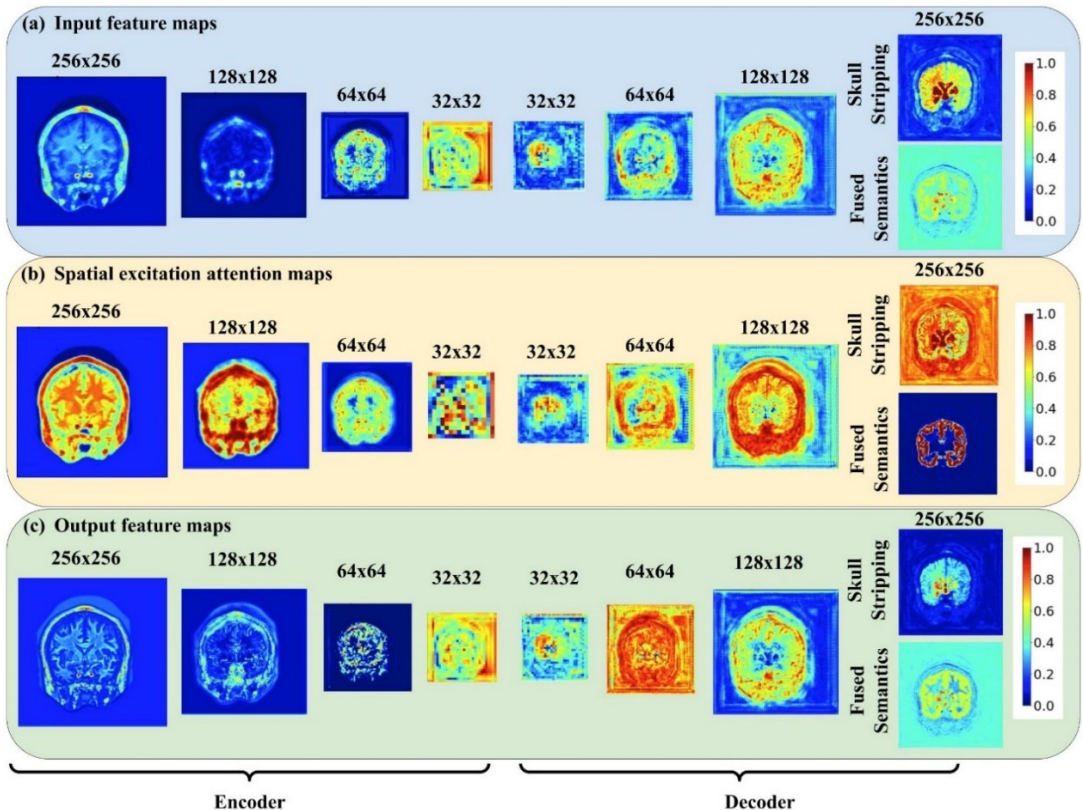

Fig 2. Visualization of (a) input feature maps, (b) spatial-wise attention maps and (c) output feature maps learned from encoders and decoders of ACEnet for segmenting coarse-grained brain structures on the 2012 MALC data set. Intensity values of all the feature maps and spatial attention maps were normalized into the range of 0 to 1. The first input features were the input image slices.

Table 2 summarizes segmentation performance of deep learning models built with the anatomical context encoding module in conjunction with different settings of the spatial context encoding module and the skull stripping module. These results indicated that the combination of the anatomical context encoding module, the spatial context encoding module, and the skull stripping module achieved the best segmentation performance on SchizBull 2008, 2012 MALC (133 structures), and Mindboggle-101 data sets. The parallel encoders with inputs of single 2D image slice and a stack of 2D image slices could further improve the segmentation performance on the dataset of 2012 MALC (27 structures) and achieved the best performance in conjunction with the skull stripping module. However, the parallel encoders did not improve the fine-grained brain structure segmentation.

Table 2. Dice scores of ACEnet with different modules on three benchmark datasets. $s$ is the number of consecutive 2D image slices stacked on top and bottom of the center slice. s=0 & s=5 indicates the presence of two parallel encoders with inputs of a single slice and a stack of multiple slices, respectively.

| Datasets | $s$=0 | $s$=0 with skull stripping | $s$=5 | $s$=5 with skull stripping | $s$=0 & $s$=5 | $s$=0 & $s$=5 with skull stripping |
|---|---|---|---|---|---|---|
| 2012 MALC (27 structures) | 0.887±0.065 | 0.888±0.062 | 0.885±0.065 | 0.885±0.065 | 0.888±0.066 | 0.890±0.062 |
| SchizBull 2008 | 0.867±0.093 | 0.870±0.092 | 0.872±0.090 | 0.872±0.089 | 0.869±0.092 | 0.872±0.092 |
| 2012 MALC (133 structures) | 0.734±0.159 | 0.739±0.148 | 0.737±0.164 | 0.746±0.143 | 0.742±0.146 | 0.743±0.143 |
| Mindboggle-101 | 0.792±0.079 | 0.799±0.078 | 0.815±0.075 | 0.820±0.076 | 0.795±0.077 | 0.797±0.077 |

To investigate how the parameter $s$ in the spatial context encoding module affects the segmentation performance, we evaluated deep learning models built with different values of $s$ using the end-to-end training setting with the presence of the anatomical context encoding module and the skull stripping module. As summarized in Table 3, the best performance for both the coarse-grained segmentation and fine-grained segmentation on the 2012 MALC dataset were achieved with $s = 5$. This value was adopted in all following experiments for the coarse-grained and fine-grained segmentation studies.

Table 3. Segmentation performance (mean ± standard deviation of Dice Score) of our method on the 2012 MALC testing dataset with different values of slice number $s$ in the spatial context encoding module.

|  | 2012 MALC (27 structures) | 2012 MALC (133 structures) |
|---|---|---|
| S=1 | 0.885±0.063 | 0.741±0.148 |
| S=3 | 0.885±0.069 | 0.743±0.145 |
| S=5 | 0.885±0.065 | 0.746±0.143 |
| S=7 | 0.883±0.080 | 0.741±0.149 |
| S=9 | 0.884±0.068 | 0.744±0.147 |

Table 4. Segmentation performance (mean ± standard deviation of Dice Score) of our methods with different training strategies.

| Datasets | First Stage | End-to-End | Two Stages |
|---|---|---|---|
| 2012 MALC (27 structures) | 0.885±0.065 | 0.885±0.065 | 0.891±0.057 |
| SchizBull 2008 | 0.872±0.090 | 0.872±0.089 | 0.881±0.074 |
| 2012 MALC (133 structures) | 0.737±0.164 | 0.746±0.143 | 0.762±0.136 |
| Mindboggle-101 | 0.815±0.075 | 0.820±0.076 | 0.825±0.074 |

*B. Ablation Study on Training Strategies*

Table 4 summarizes segmentation performance of the deep learning models trained using different training strategies. These results indicated that the end-to-end model yielded better results than the model without the skull stripping module obtained in the first stage, and the model obtained in the second stage obtained the best performance. We adopted the two-stage training strategy in all following experiments.

*C. Model Complexity*

We compared model complexity between baseline (Roy et al., 2018) and models with our proposed modules (all included the Context Encoding Module) based on images of $256 \times 256$. As summarized in Table 5, the baseline model with an input of single image slice had $3.551 \times 10^6$ parameters, and the Context Encoding Module added $4.38 \times 10^5$ (an increase of 12.3%) parameters to the baseline model. Since the skull stripping module shares the first three decoders with the backbone's decoders, it added $2 \times 10^4$ (an increase of 0.05%) parameters to a model with the Context Encoding Module.

An input of the stacked image volumes ($s = 5$) had $1.52 \times 10^5$ more (an increase of 3.8%) parameters than the input of single image slice ($s = 0$). The parallel encoders increase the model complexity substantially, with an increase of 41.7% and 36.5% in the number of parameters compared with the models with $s = 0$ and $s = 5$ respectively. However, their segmentation performance did not increase with the number of parameters, except on the 2012 MALC dataset for segmenting coarse-grained brain structures, as indicated by the results summarized in Table 2.

Overall, the model, with the anatomical context encoding module, the skull stripping module, and the spatial context encoding module (a stack of image slices with $s = 5$) obtained the best segmentation performance at a computation cost of 16.6% increase in the number of parameters compared with the baseline model.

Table 5. Model complexity. $S$ is the number of consecutive 2D image slices stacked on top and bottom of the center slice. & indicates the presence of two parallel encoders which take both inputs of a single slice and a stack of 2.5D stack of images.

| Models | baseline | s=0 | s=0 with skull stripping | s=5 | s=5 with skull stripping | s=0 & s=5 | s=0 & s=5 with skull stripping |
|---|---|---|---|---|---|---|---|
| Number of parameters | $3.551 \times 10^6$ | $3.989 \times 10^6$ | $3.991 \times 10^6$ | $4.141 \times 10^6$ | $4.142 \times 10^6$ | $5.653 \times 10^6$ | $5.655 \times 10^6$ |

*D. Comparison with Competing Methods for Segmenting Coarse-grained Brain Structures*

Tables 6 and 7 summarize segmentation performance obtained by competing methods under comparison for segmenting coarse-grained brain structures on the 2012 MALC dataset and the SchizBull 2008 dataset, respectively. As summarized in Table 6, ACEnet obtained a mean Dice Score of 0.891, an improvement of 1.7% compared with the second best method, i.e., QuickNAT V2. The data augmentation further improved our method though the improvement was moderated. As summarized in Table 7, ACEnet also obtained the best segmentation performance on the SchizBull 2008 dataset with an improvement of 2.2% compared with the second-best method, i.e., QuickNAT V2. Interestingly, the methods built upon 2D CNNs obtained better performance than those build upon 3D CNNs for segmenting coarse-grained brain structures.

Table 6. Comparison of deep learning methods for segmenting coarse-grained brain structures based on the 2012 MALC testing dataset, including segmentation accuracy measured by Dice score and the number of parameters in each model. − indicates parameters are not reported in their respective papers, † indicates segmentation performance obtained from their respective papers, and * indicates a model trained with data augmentation.

| Methods | CNNs | Parameters | Dice Score (mean ± standard deviation) |
|---|---|---|---|
| 3D U-Net (Çiçek et al., 2016) | 3D | $6.444 \times 10^6$ | 0.859±0.082 |
| SLANT8 (Huo et al., 2019) † | 3D | − | 0.817±0.036 |
| SLANT27 (Huo et al., 2019) † | 3D | − | 0.823±0.037 |
| MO-Net (Dai et al., 2019) † | 3D | − | 0.838±0.049 |
| 3DQuantized-Unet (Paschali et al., 2019) † | 3D | $2.0 \times 10^6$ | 0.844±0.006 |
| DeepNAT (Wachinger et al., 2018) † | 3D | $2.7 \times 10^6$ | 0.894 |
| SD-Net (Roy et al., 2017) † | 2D | − | 0.850±0.080 |
| SD-Net (Roy et al., 2017) | 2D | $5.7 \times 10^5$ | 0.860±0.097 |
| U-Net (Ronneberger et al., 2015) | 2D | $5.178 \times 10^6$ | 0.869±0.080 |
| QuickNAT (Roy et al., 2019) | 2D | $3.551 \times 10^6$ | 0.874±0.067 |
| QuickNAT V2 (Roy et al., 2018) | 2D | $3.551 \times 10^6$ | 0.876±0.067 |
| ACEnet | 2D | $4.142 \times 10^6$ | 0.891±0.057 |
| ACEnet* | 2D | $4.142 \times 10^6$ | 0.897±0.057 |

Table 7. Comparison of deep learning methods for segmenting coarse-grained brain structures on the SchizBull 2008 testing dataset.

| Methods | CNNs | Parameters | Dice Score (mean ± standard deviation) |
|---|---|---|---|
| U-Net (Çiçek et al., 2016) | 3D | $6.444 \times 10^6$ | 0.857±0.097 |
| SD-Net (Roy et al., 2017) | 2D | $5.7 \times 10^5$ | 0.856±0.098 |
| U-Net (Ronneberger et al., 2015) | 2D | $5.178 \times 10^6$ | 0.862±0.096 |
| QuickNAT V2 (Roy et al., 2018) | 2D | $3.551 \times 10^6$ | 0.862±0.095 |
| ACEnet | 2D | $4.142 \times 10^6$ | 0.881±0.074 |

Table 8 summarizes image segmentation performance measured by Dice Score, Jaccard Index, and Hausdorff Distance obtained by the top-two deep learning models, i.e., QuickNAT V2 and ACEnet on both the 2012 MALC dataset with 27 structures and the SchizBull 2008 dataset. These results demonstrated that ACEnet performed consistently better than QuickNAT V2 in terms of Dice score, Jaccard score, and Hausdorff Distance. The results of skull stripping were promising with Dice scores greater than 0.987.

Table 8. Segmentation performance (mean ± standard deviation) of our methods and QuickNAT V2 on two coarse-grained benchmark datasets. Skull Stripping is reported on Mean Dice Score for our model on testing data.

| Datasets | Performance measures | QuickNAT V2 | ACEnet |
|---|---|---|---|
| MALC (27 structures) | Dice | 0.876±0.077 | 0.891±0.057 |
| | Jaccard | 0.777±0.122 | 0.809±0.088 |
| | Skull-stripping (Dice) | -- | 0.987±0.012 |
| | Hausdorff Distance | 4.156±0.620 | 3.965±0.553 |
| SchizBull 2008 | Dice (test) | 0.862±0.095 | 0.881±0.074 |
| | Dice (validation) | 0.862±0.084 | 0.880±0.087 |
| | Jaccard | 0.766±0.131 | 0.796±0.122 |
| | Skull-stripping (Dice) | -- | 0.993±0.006 |
| | Hausdorff Distance | 4.347±0.453 | 4.150±0.413 |

Representative segmentation results are visualized in Fig. 3 with zoomed-in regions to highlight differences among results obtained by the methods under comparison. As illustrated by the results on the left column, ACEnet obtained visually better segmentation results than QuickNAT V2 for segmenting the left lateral ventricle on the MALC dataset. The results shown on the second left column indicated that our method had better performance than QuickNAT V2 for segmenting bilateral amygdala on the SchiBull 2008 dataset.

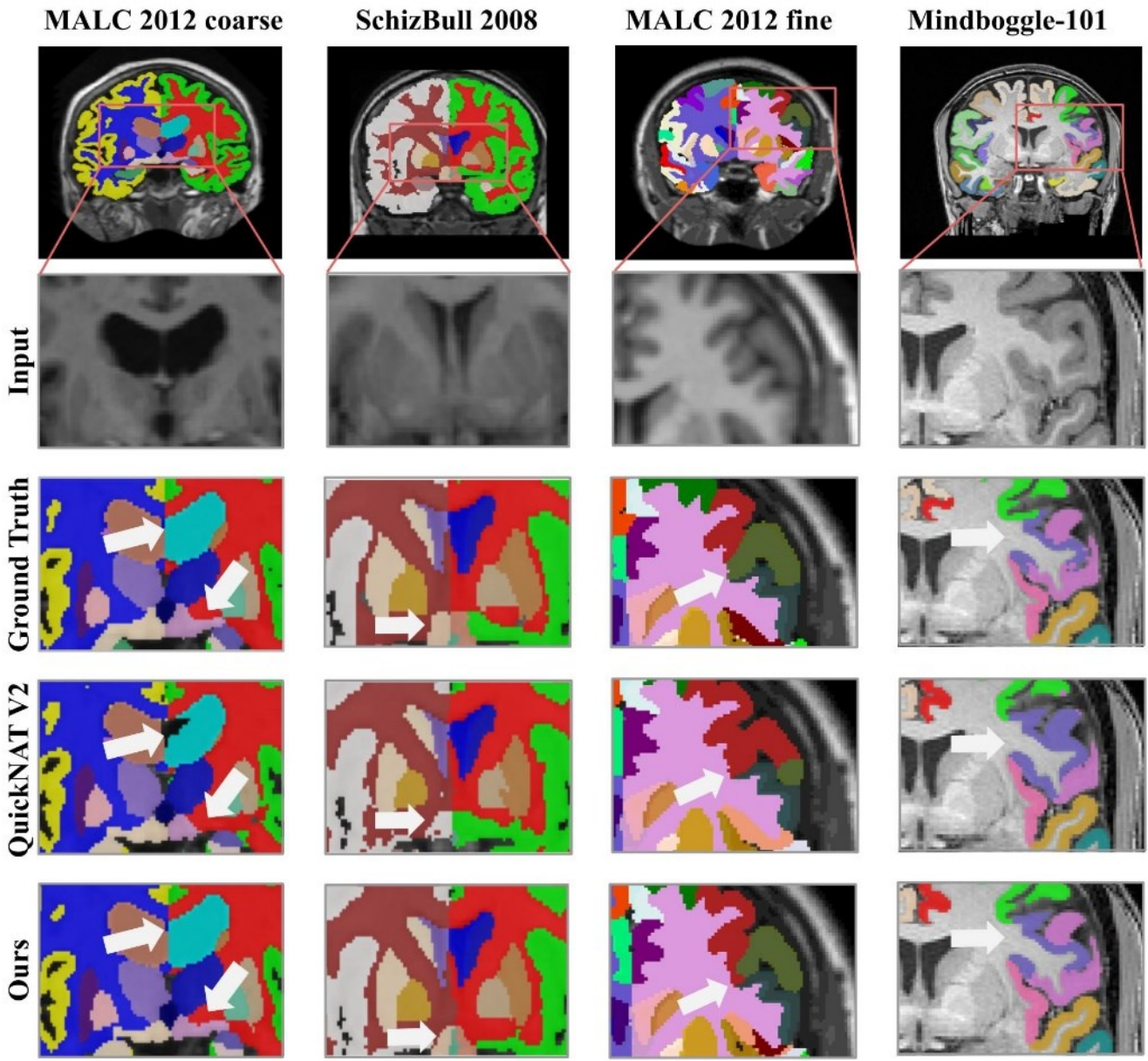

Fig. 3. Representative input image slices, ground truth, and outputs of QuickNAT V2 and ACEnet for segmenting coarse-grained brain structures on the 2012 MALC dataset and the SchiBull 2008 dataset (left two columns) and fine-grained brain structures on the 2012 MALC dataset and the Mindboggle 101 dataset (right two columns), respectively.

As shown in Fig. 4 and Fig. 5, statistical comparisons on Dice Scores of individual structures also indicated that our method had significantly better performance than QuickNAT V2 for segmenting most brain structures on the 2012 MALC dataset and SchiBull dataset ($p$<0.05, two-sided Wilcoxon signed rank test). Overall, two-sided Wilcoxon signed rank tests indicated that our method performed significantly better than QuickNAT V2 for segmenting the coarse-grained brain structures in terms of Dice score on both the MALC and SchiBull datasets with $p$ values of $5.61 \times 10^{-6}$ and $7.95 \times 10^{-7}$, respectively.

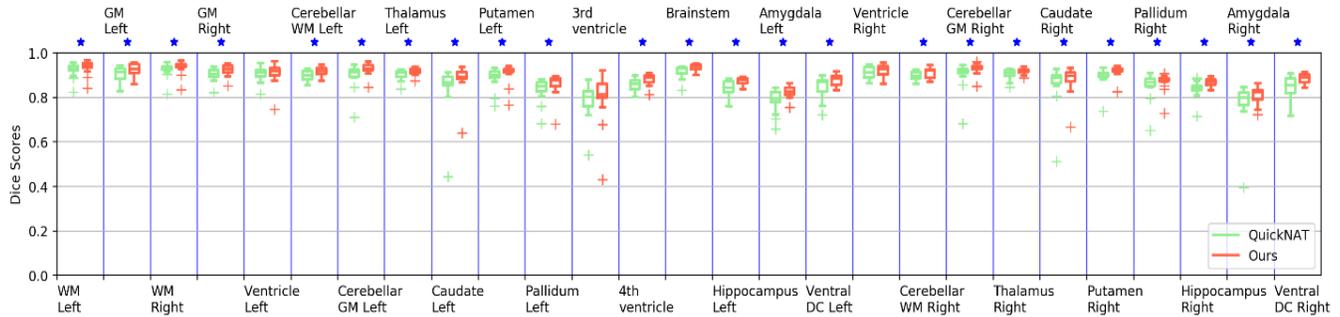

Fig. 4. Box plot of Dice scores of 27 structures obtained by ACEnet (ours) and QuickNAT V2 on the 2012 MALC coarse-grained structure dataset with 15 TI MRI test scans. WM indicates White Matter and GM indicates Grey Matter. The star (⋆) symbol represents the statistical significance (p ≤ 0.05, two-side Wilcoxon signed rank test).

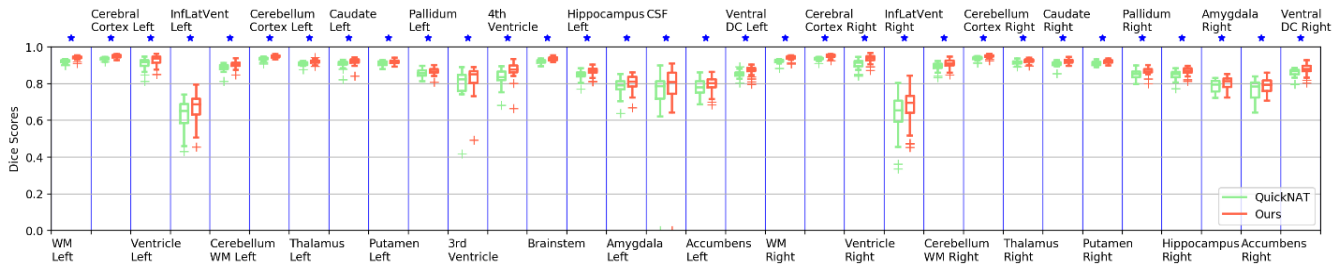

Fig. 5. Box plot of Dice scores of 32 structures obtained by ACEnet (ours) and QuickNAT V2 on the SchizBull 2008 coarse-grained structure dataset with 30 TI MRI test scans. WM indicates White Matter and GM indicates Grey Matter. The star (⋆) symbol represents the statistical significance (p ≤ 0.05, two-side Wilcoxon signed rank test).

Table 9. Comparison of deep learning methods for segmenting coarse-grained brain structures based on the 2012 MALC testing dataset, including segmentation accuracy measured by Dice score and the number of parameters in each model. − indicates parameters are not reported from their respective papers, † indicates segmentation performance obtained from their respective papers, and * indicates a model trained with data augmentation.

| Methods | CNNs | Parameters | Dice Score (mean $\pm$ standard deviation) |
| --- | --- | --- | --- |
| 3D U-Net (Çiçek et al., 2016) | 3D | $7.687 \times 10^6$ | 0.704±0.156 |
| SLANT8 (Huo et al., 2019) † | 3D | − | 0.768±0.011 |
| SLANT27 (Huo et al., 2019) † | 3D | − | 0.776±0.011 |
| Seg-Net (de Brebisson and Montana, 2015) † | 3D | $3.056 \times 10^7$ | 0.725±0.163 |
| SD-Net (Roy et al., 2017) | 2D | $5.7 \times 10^5$ | 0.628±0.205 |
| 2D U-Net (Ronneberger et al., 2015) | 2D | $5.178 \times 10^6$ | 0.688±0.156 |
| QuickNAT V2 (Roy et al., 2018) | 2D | $3.551 \times 10^6$ | 0.689±0.161 |
| ACEnet | 2D | $4.142 \times 10^6$ | 0.762±0.136 |
| ACEnet* | 2D | $4.142 \times 10^6$ | 0.771±0.134 |

*E. Comparison with Alternative Methods with 2D or 3D CNNs for the Fine-grained Segmentation*

Tables 9 and 10 summarize segmentation performance obtained by competing methods under comparison for segmenting fine-grained brain structures on the 2012 MALC dataset and the Mindboggle-101 dataset, respectively. As summarized in Table 9, ACEnet obtained a mean Dice Score of 0.762, an improvement of 9.6% compared with the second best method with 2D CNNs, i.e., QuickNAT V2. The data augmentation further improved our method and achieved segmentation accuracy close to those obtained by the methods built upon 3D CNNs. It is worth noting that the best model, i.e., SLANT 27, was trained on a larger training dataset and a larger augmentation dataset. As summarized in Table 10, ACEnet obtained the best segmentation performance on the Mindboggle-101 dataset with a Dice score of 82.5% and an improvement of 4.2% compared with the second-best method, i.e., 3D U-Net, and an improvement of 5.8% compared with QuickNAT V2.

Table 10. Comparison of deep learning methods for segmenting fine-grained brain structures on the Mindboggle-101 testing dataset.

| Methods | CNNs | Parameters | Dice Score (mean ± standard deviation) |
|---|---|---|---|
| U-Net (Çiçek et al., 2016) | 3D | $7.687 \times 10^6$ | 0.790±0.079 |
| SD-Net (Roy et al., 2017) | 2D | $5.7 \times 10^5$ | 0.754±0.089 |
| U-Net (Ronneberger et al., 2015) | 2D | $5.178 \times 10^6$ | 0.767±0.086 |
| QuickNAT V2 (Roy et al., 2018) | 2D | $3.551 \times 10^6$ | 0.777±0.082 |
| ACEnet | 2D | $4.142 \times 10^6$ | 0.825±0.074 |

Table 11 summarizes image segmentation performance measured by Dice Score, Jaccard Index, and Hausdorff Distance obtained by the top-two deep learning models built upon 2D CNNs, i.e., QuickNAT V2 and ACEnet on both the 2012 MALC dataset with 133 structures and the Mindboggle-101 dataset. These results demonstrated that ACEnet performed consistently better than QuickNAT V2 in terms of Dice score, Jaccard score, and Hausdorff Distance. Specifically, ACEnet obtained an improvement of 9.6% and 5.8% compared with QuickNAT V2 in terms Dice score on the 2012 MALC dataset with 133 structures and the Mindboggle-101 dataset, respectively. The results of skull stripping were promising too with Dice scores greater than 0.976.

Table 11. Segmentation performance (mean ± standard deviation) of our methods and QuickNAT V2 on two fine-grained benchmark datasets. Skull Stripping is reported on Mean Dice Score for our model.

| Datasets | Performance measures | QuickNAT V2 | ACEnet |
|---|---|---|---|
| MALC (133 structures) | Dice | 0.689±0.161 | 0.762±0.136 |
| | Jaccard | 0.547±0.176 | 0.633±0.162 |
| | Skull-stripping (Dice) | -- | 0.987±0.014 |
| | Hausdorff Distance | 6.682±0.614 | 5.794±0.387 |
| Mindboggle-101 | Dice (test) | 0.777±0.082 | 0.825±0.074 |
| | Dice (validation) | 0.763±0.103 | 0.804±0.100 |
| | Jaccard | 0.643±0.107 | 0.704±0.101 |
| | Skull-stripping (Dice) | -- | 0.976±0.024 |
| | Hausdorff Distance | 6.523±0.382 | 6.454±0.456 |

Representative segmentation results for segmenting the fine-grained brain structures are visualized in Fig. 3 (right two columns) with zoomed-in regions to highlight differences among results obtained by the methods under comparison, indicating that ACEnet obtained visually better segmentation results than QuickNAT V2 for segmenting cortical areas on both the 2012 MALC dataset and the Mindboggle-101 dataset. As illustrated in Fig. 6 and Fig. 7, statistical comparisons on Dice scores of individual structures have also indicated that our method had significantly better performance than QuickNAT V2 for segmenting most of the brain structures on both the 2012 MALC and Mindboggle-101 datasets ($p<0.05$, two-side Wilcoxon signed rank test). Overall, two-side Wilcoxon signed rank tests indicated that our method performed significantly better than QuickNAT V2 for segmenting the fine-grained brain structures in terms of Dice score on both the MALC and Mindboggle-101 datasets with $p$ values of $3.22 \times 10^{-24}$ and $7.58 \times 10^{-12}$, respectively.

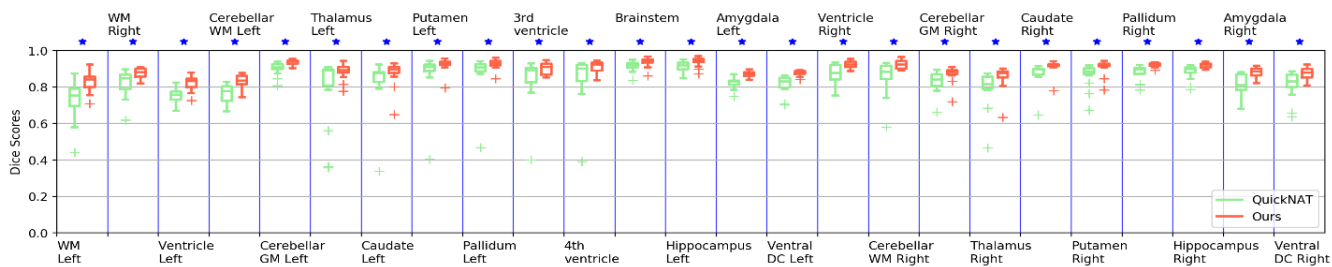

Fig. 6. Box plot of Dice scores of 133 structures obtained by ACEnet (ours) and QuickNAT V2 on the 2012 MALC fine-grained structure dataset with 15 T1 MRI test scans. In this plot we show 25 subcortical structures for visualization. WM indicates White Matter and GM indicates Grey Matter. The star (⋆) symbol represents the statistical significance (p ≤ 0.05, two-side Wilcoxon signed rank test).

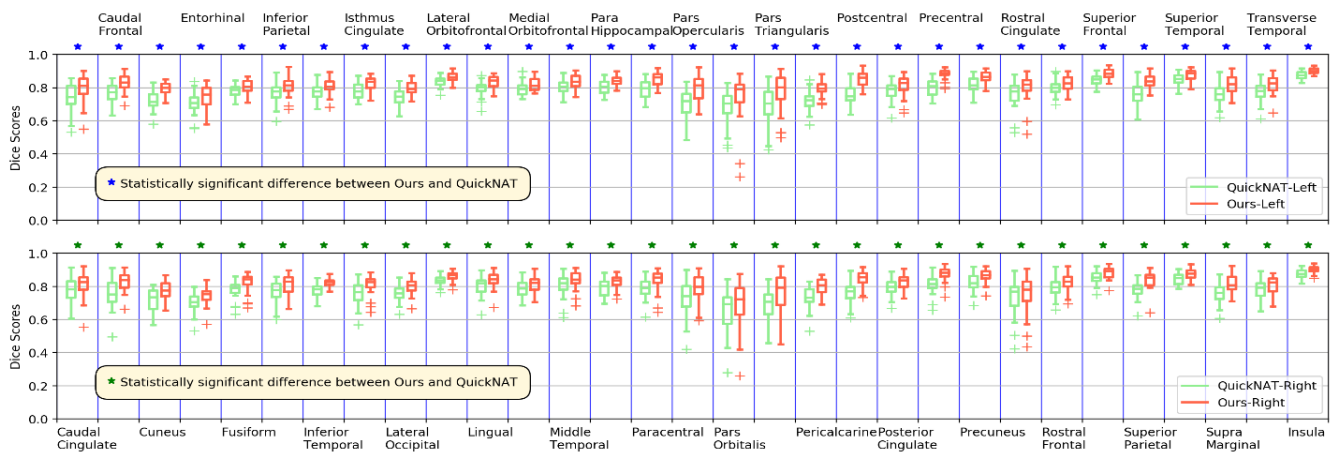

Fig. 7. Box plot of Dice scores of 62 structures obtained by ACEnet (ours) and QuickNAT V2 on Mindboggle-101 fine-grained structure dataset with 30 T1 MRI test scans. The top and bottom plots show the segmentation performance on structures of the left and right hemispheres, respectively. The star (⋆) symbol represents the statistical significance (p ≤ 0.05, two-side Wilcoxon signed rank test).

## DISCUSSIONS

We propose a new deep learning method, Anatomy Context-Encoding network (ACEnet), to segment brain structures from 3D MRI head scans using 2D CNNs enhanced by 3D spatial and anatomical context information. Experimental results based on three benchmark datasets have demonstrated that our method could achieve better segmentation accuracy than state-of-the-art alternative deep learning methods for segmenting coarse-grained brain structures and comparable performance for segmentation fine-grained brain structures. Furthermore, the skull stripping module and the two-stage training strategy also obtained promising performance. The deep learning segmentation models built by our method could segment an MRI head scan of 256×256×256 within ~9 seconds on a NVIDIA TITAN XP GPU, facilitating real-time applications.

We have compared our method with state-of-the-art brain image segmentation methods built upon 2D CNNs and 3D CNNs with a focus on those built upon 2D CNNs for computational efficiency. Particularly, we directly compared our method with SD-net (Roy et al., 2017), 2D Unet (Ronneberger et al., 2015), QuickNAT V2 (Roy et al., 2018), and 3D Unet (Çiçek et al., 2016). We evaluated these methods based on 3 publicly available datasets, including the 2012 MALC dataset with 27 and 133 brain structures, the Mindboggle dataset, and the SchizBull dataset. Based on these datasets, we evaluated the competing deep learning methods for segmenting coarse-grained and find-grained brain structures, respectively. The 2012 MALC dataset also provides training and testing lists to facilitate comparisons among image segmentation methods evaluated based on the same training and testing lists. Based on the 2012 MALC dataset we also indirectly compared our method with MO-Net (Dai et al., 2019), Seg-Net (de Brebisson and Montana, 2015), SLANT (Huo et al., 2019), DeepNAT (Wachinger et al., 2018), and 3DQuantized-Unet (Paschali et al., 2019). Comparison results summarized in Tables 6 and 7 demonstrated that ACEnet obtained the best segmentation performance among all methods under comparison, including those build upon 3D CNNs, for segmenting coarse-grained brain structures. Comparison results summarized in Tables 9 and 10 demonstrated that ACEnet obtained promising performance, better than those obtained by the alternative methods built upon 2D CNNs and comparable to those obtained by the methods built upon 3D CNNs, such as SLANT 27. However, ACEnet is computationally more efficient than SLANT27 that was trained on a larger training dataset.

Our method is built upon QuickNAT V2 with three proposed modules. First, our method has a spatial context encoding module to encode 3D spatial context information of consecutive image slices as a multi-channel input. This module uses 2D convolutional layers to extract 3D spatial context information for computational efficiency. Ablation studies indicated that this module could improve the segmentation performance for both the coarse-grained and fine-grained brain structure segmentation tasks, supported by quantitative evaluation results summarized in Tables 1, 2, 3, and 4 and visualization results shown in Fig. 3.

Second, our method has an anatomical context encoding module to guide 2D CNNs to focus on brain structures present in the center image slices under consideration. This module consists of an attention factor to encode the anatomical information, learned by optimizing an anatomical context encoding classification loss to identify the presence of specific brain structures in the center image slices. This anatomical context encoding module improves the brain structure segmentation in two aspects. First, the anatomical context information acts as an attention factor that provides a global anatomical prior to squeeze the intensity ambiguity between structures with similar appearances. Different from training separate CNNs for segmenting different brain structures (Huo et al., 2019), the attention factor facilitates a single segmentation model to adaptively encode anatomical information for individual image slices. Second, the anatomical context information also serves as a regularizer to guide the 2D CNNs to focus on brain structures present in the center image slices under consideration, rather than all brain structures to be segmented. Such a regularizer could potentially make the segmentation more robust, especially for the fine-grained brain structure segmentation as only a small number of brain structure are present in individual image slices and therefore yield a classification problem with unbalanced training samples. The ablation studies in conjunction with the representative spatial-wise attention maps and image feature maps before and after modulation by the sc-SE blocks shown in Fig. 2 all indicated that the sc-SE blocks and the anatomical context encoding module effectively improved the image segmentation performance.

Finally, our method has a skull stripping module as an auxiliary task to guide 2D CNNs to focus on brain structures rather than non-brain structures. The ablation studies indicated that this skull-stripping module could improve the brain structure segmentation performance no matter whether the end-to-end or the two-stage training strategies was used to training the segmentation. The experimental results also indicated that the two-stage training strategy could improve the segmentation results compared with the end-to-end training, consistent with findings in prior studies (Ren et al., 2015).

The present study has following limitations. First, we did not tune the hyperparameters of the proposed method exhaustively due to high computational cost. Instead, we tuned the hyperparameters by fixing some of them, which may lead to inferior performance. Second, we used simple data augmentation method to augment the training data. The results of SLANT indicated that multi-atlas image segmentation can be used to augment the training data, albeit computationally expensive (Huo et al., 2019). We will adopt deep learning based image registration methods to improve the computational efficiency of multi-atlas image segmentation methods to augment the training data (Li and Fan, 2017, 2018, 2020) in our future studies. Third, we compared our method indirectly with some competing methods based on the 2012 MALC dataset. Although most of the evaluations were carried out on the same training and testing data (except SLANT), the comparison results should be interpreted with a caveat that their performance is hinged on training strategies including data argumentation.

## CONCLUSIONS

Anatomy Context-Encoding network (ACEnet) provides a computationally efficient solution for both the coarse-grained and fine-grained brain structure segmentation tasks. Our method could be potentially applied to other image segmentation studies, such as segmentation of white matter hyperintensities and brain tumors (Li et al., 2018; Zhao et al., 2018).

## ACKNOWLEDGMENT

This study was supported in part by National Institutes of Health grants [EB022573, MH120811].